\documentclass[aps,prd,onecolumn,groupedaddress,preprintnumbers,superscriptaddress,showpacs]{revtex4}
\usepackage{amsmath,amssymb,latexsym}

\newcommand{\cbar}[1]{{~#1\!\!\!\!-}}

\begin{document}

\title{
Quadrupole formula for Kaluza-Klein modes in the braneworld
}

\author{Shunichiro Kinoshita}\email{kinoshita_at_utap.phys.s.u-tokyo.ac.jp}
\author{Hideaki Kudoh}
\author{Yuuiti Sendouda}
\affiliation{Department of Physics, School of Science, University of Tokyo, 7-3-1 Hongo, Bunkyo, Tokyo 113-0033, Japan}
\author{Katsuhiko Sato}
\affiliation{Department of Physics, School of Science, University of Tokyo, 7-3-1 Hongo, Bunkyo, Tokyo 113-0033, Japan}
\affiliation{Research Center for the Early Universe (RESCEU), School of Science, University of Tokyo, 7-3-1 Hongo, Bunkyo, Tokyo 113-0033, Japan}

\preprint{UTAP-524} 
\preprint{RESCEU-7/05}

\date{\today}

\begin{abstract}
The quadrupole formula in four-dimensional Einstein gravity is a useful
 tool to describe gravitational wave radiation.
We derive the quadrupole formula for the Kaluza-Klein (KK) modes in the
 Randall-Sundrum braneworld model.
The quadrupole formula provides transparent
 representation of the exterior weak gravitational field induced by
 localized sources.
We find that a general isolated dynamical source gives rise to the
 $1/r^2$ correction to the leading $1/r$ gravitational field.
We apply the formula to an evaluation of the effective energy carried by the
 KK modes from the viewpoint of an observer on the brane.
Contrary to the ordinary gravitational waves (zero mode), the flux of
 the induced KK modes by the non-spherical part of the quadrupole moment
 vanishes at infinity and only the spherical part contributes to the flux.
Since the effect of the KK modes appears in the linear order of the
 metric perturbations, the effective energy flux observed on the brane is not always positive, but can become negative depending on the motion of the localized sources. 
\end{abstract}

\pacs{04.30.-w, 04.50.+h}
\maketitle

\section{Introduction}

Strong gravitational phenomena, such as gravitational collapse and black
hole formation, have long been intriguing subjects of General Relativity
in the context of astrophysics.
Likewise, studying strongly gravitating dynamics in frameworks of recent
models of the universe in which we live on a 4D membrane (braneworlds)
will promisingly open up new aspects of the models
themselves~\cite{BGM,GD, Neves, Casadio:2004nz}.
One of the sites is naturally gravitational collapse occurring typically
in the last stage of stellar evolution or during the early era of the
universe (the latter corresponds primordial black holes), which would
experience sufficiently high density and energy so that effects of extra
dimensions might not be negligible.

The model proposed by Randall and Sundrum (RS) \cite{RS1,RS2} is a
simple, but successful model that realizes a braneworld.
So far no serious drawback to the model has been known. 
 (See, e.g., \cite{Maartens} for a comprehensive review). 
In the model, perturbative analysis of weak gravity reveals that the KK
modes give the correction ($\propto r^{-3}$) to the Newton potential
($\propto r^{-1}$) for static isolated
sources~\cite{RS2,Garriga:1999yh,GKR}.
Furthermore it is known that the
gravity induced by relativistic stars \cite{Wiseman:2001xt} and black
holes localized on the
brane~\cite{Kudoh:2003xz,Kudoh:2003vg,Kudoh:2004kf} is well described by
the {\it lower dimensional} 4D Einstein gravity on the brane.
The induced gravity in the latter case is approximated by the black
string when the mass is sufficiently large, although the subject of the
black hole in the braneworld is a little
controversial~\cite{Tanaka:2002rb,Emparan:2002px} (see also,
e.g. \cite{Anderson:2004md}).

Our knowledge of strongly gravitating {\it dynamical} processes in the
braneworld is, however, very limited.
Even for a simple spherically symmetric collapse, the external
gravitational field including Kaluza-Klein (KK) modes excited during the
process has not been fully investigated.
In 4D General Relativity, as a result of Birkhoff's theorem, all
spherically symmetric vacuum spacetimes become the Schwarzschild
solution even implying that any spherical collapse cannot radiate
gravitational waves;
only non-spherical dynamical processes produce gravitational waves.
Their amplitude and energy loss rate are well described by the so-called
quadrupole formula,
whose prediction for the energy loss rate in the binary pulsars, e.g.
PSR1913+16 and PSR1534+12, precisely agree with the observed orbital
decay rate~\cite{Taylor}.
Thus the quadrupole formula is of interest to test alternative
theories of gravity~\cite{Will:1993}.
Even in higher-dimensional spacetimes, as far as we consider spherically
symmetric collapse, the situation does not 
change~\cite{Gibbons:2003}.
The higher-dimensional quadrupole formula in the asymptotically flat spacetimes has been recently discussed
in~\cite{Lemos} (for even $D$-dimensions).
However, in realistic braneworld models such as RS, matter fields are confined on the brane and seemingly spherical collapse of such fields 
is not really so from the viewpoint of the bulk so that the Kaluza-Klein
modes would be inevitably induced on the brane and the non-local
effect from the bulk gravity complicates the problem.  
In \cite{BGM}, considering homogeneous spherical dust collapse on
the brane it was shown that outside the dust sphere the induced metric
could not remain static and vacuum.
Hence it is necessary and important to investigate the dynamical process
including dynamically generated  KK gravitons,
it will allow insight into the BH formation and astrophysical implications of the
braneworld~\cite{Emparan:2002jp,Clancy:2003zd,Sendouda:2003dc,Sendouda:2004hz}.
(See e.g. \cite{Seahra:2004fg} for gravity wave perturbations using the black
string.)

The purpose of this paper is to have insight into physically interesting dynamical processes in the braneworld.
Toward this direction, we investigate a gravitational field induced by an isolated dynamical system on the brane.
We employ the linear perturbation theory of the RS
model~\cite{RS2,Garriga:1999yh,GKR} to accomplish a self-contained
analysis.
We first derive a quadrupole formula which describes the (KK) gravity outside matter distribution by means of quadrupole moments.
By applying this formula, we can successfully describe weak
gravity far away from a source once quadrupole
moments of an isolated system are determined.
Based on the quadrupole representation of the induced metric, we also
discuss effective energy that is transferred by the KK modes.  
This kind of energy flux corresponds in the linear order to the flux of so-called dark radiation in the braneworld cosmology~\cite{Mukohyama:2000}. 
We find that the KK modes contribute to the flux only through the
projected bulk Weyl tensor as an effective matter, but not through
energy-momentum tensor of the gravitational field itself. 
Moreover, this effective energy density is not always positive, and in fact it becomes negative in the case of the homogeneous dust collapse.

The paper is organized as follows. In Sec.~\ref{sec:Linear perturbation}, we review the linear perturbation theory of the RS model discussing the Green's function for the dynamical source, and derive the quadrupole formula for dynamical systems.
In Sec.~\ref{sec:Applications}, we apply the formulation to a spherically symmetric system on the brane. As a concrete example, we discuss spherical dust collapse and estimate the change rate of the effective energy associated with the KK modes. General non-spherically symmetric cases are also discussed in this section.

\section{Linear perturbation in the braneworld}
\label{sec:Linear perturbation}

\subsection{Green's function}

Let us begin with metric perturbation in the RS model in order to discuss weak gravitational field generated by an isolated dynamical system on the brane.
We follow the procedure given in~\cite{Garriga:1999yh}.
(Notation in this paper is the same as in~\cite{Garriga:1999yh}.)
Denoting metric perturbation in the RS gauge as $h_{\mu\nu}$, the 5D metric is given by
\begin{equation}
\mathrm d s^2 = \frac{\ell^2}{z^2}\left[
 dz^2 + \eta_{\mu\nu}  
  \mathrm dx^\mu\mathrm dx^\nu \right] 
 +  h_{\mu\nu}\mathrm dx^\mu\mathrm dx^\nu ,
\end{equation}
where $\eta_{\mu\nu}$ is the 4D Minkowski metric and $\ell$ is the curvature radius of anti-de Sitter (AdS).
In this gauge linearized 5D Einstein's equation becomes
\begin{equation}
\left[{}^{(4)}\square + \frac{\partial^2}{\partial z^2} +
 \frac{1}{z}\frac{\partial}{\partial z} - \frac{4}{z^2}\right]h_{\mu\nu}
= -2\kappa \Sigma_{\mu\nu}\frac{\ell}{z}\delta(z-\ell),\label{linearized}
\end{equation}
where the source term is given by
\begin{equation}
\Sigma_{\mu\nu} = \left(T_{\mu\nu} - \frac{1}{3}\gamma_{\mu\nu}T\right)
 + 2\kappa^{-1}\hat\xi^5{}_{,\mu\nu}, \quad \gamma_{\mu\nu} \equiv
 \frac{\ell^2}{z^2} \eta_{\mu\nu}.
\end{equation}
Here $T_{\mu\nu}$ is the energy-momentum tensor of matter on the brane at
$z=\ell$ and
$\hat\xi^5$ expresses location of the brane in the RS gauge, which is related to the matter source on the brane~\cite{Garriga:1999yh}.  
4D Newton's constant is $G_4$ and we use $\kappa = 8\pi G_4\ell$.

Using retarded Green's function the formal solution of
Eq.~(\ref{linearized}) is obtained as
\begin{equation}
h_{\mu\nu} = -2\kappa\int\mathrm d^4x'
 G(x,z,x',\ell)\Sigma_{\mu\nu}(x'),\label{solution}
\end{equation}
where $G(x,z,x',z')$ is decomposed into two parts;
\begin{equation}
G(x,z,x',z') = G_\mathrm{zero} + G_\mathrm{KK}.
\end{equation}
$G_\mathrm{zero}$ is the usual Green's function of massless scalar fields and
$G_\mathrm{KK}$ is that of massive Kaluza-Klein modes.
Correspondingly perturbed metric is written as 
$h_{\mu\nu} = h_{\mu\nu}^\mathrm{zero} + h_{\mu\nu}^\mathrm{KK}$,
where we note that for the RS single-brane model the zero mode
corresponds to the linear perturbation of 4D Einstein
gravity~\cite{Garriga:1999yh}.
The Fourier components of the KK modes are given by
\begin{equation}
\widehat G_{\mathrm{KK}}(\omega;\mathbf x,z,\mathbf x',\ell) =
 -\int^\infty_0\mathrm dm\int\frac{\mathrm
 d^3\mathbf k}{(2\pi)^3} \frac{u_m(z)u_m(\ell)}{\mathbf k^2-\omega^2+m^2}
e^{i\mathbf k\cdot(\mathbf x-\mathbf x')},
\end{equation}
where the mode function is 
$
u_m(z)=\sqrt\frac{m\ell}{2}\frac{J_1(m\ell)Y_2(mz)-Y_1(m\ell)J_2(mz)}
{\sqrt{J_1(m\ell)^2+Y_1(m\ell)^2}}.
$
In this paper, hat denotes Fourier components.

We are interested in the case that matter is non-relativistic and its density is low, hence the induced gravity is well described by linear perturbations. 
Then only low-frequency modes $\omega\ll 1/\ell$ are relevant.
(The $m\ell \gg 1$ case was investigated in~\cite{Sasaki:1999mi}.)
Expanding the mode function about $m\ell \ll 1$, one gets
\begin{equation}
\begin{aligned}
\widehat G_{\mathrm{KK}}(\omega;\mathbf x,z,\mathbf x',\ell)
&= - \int^\infty_0\mathrm dm
 \frac{m\ell}{2}\left(\frac{\ell}{z}\right)^2\int\frac{\mathrm d^3\mathbf
 k}{(2\pi)^3} \frac{e^{i\mathbf
 k\cdot(\mathbf x-\mathbf x')}}{\mathbf k^2-\omega^2+m^2}+ \mathcal
 O(m^3\ell^3),
\\
&= -\frac{1}{4\pi
 R}\int^\infty_0\mathrm dm\frac{m\ell}{2}\left(\frac{\ell}{z}\right)^2
 e^{iR\sqrt{\omega^2-m^2}}+ \mathcal O(m^3\ell^3), 
\\
&\simeq \frac{\ell^3}{8\pi z^2}\left(\frac{i\omega}{R^2}-\frac{1}{R^3}\right)
 e^{i\omega R},\label{KK Green's function}
\end{aligned}
\end{equation}
where 
$R=|\mathbf x-\mathbf x'|$.
Here we are interested in the gravitational field on the brane so that
we have set $z\simeq \ell$.
Note that very massive modes with $m > \omega$ are exponentially
suppressed by the factor $e^{-R\sqrt{m^2 - \omega^2}}$ in the region far
from the source, while only the KK modes with $m<\omega$ contribute to the $m$ integral.
In the static case ($\omega = 0$) the first term vanishes and the second gives the well-known KK correction to the Newtonian potential, which is proportional to $\ell^2/R^3$ \cite{RS2,Garriga:1999yh,GKR}.  
In the dynamical case ($\omega \neq 0$), however, the first term survives. 
This term corresponds to dynamical KK modes.
This kind of difference of falloff between dynamical and static terms is
known to occur in higher dimensional Green's functions~\cite{Lemos}, whereas for the 4D massless Green's function both the static potential and the wave form (dynamical term) fall as $r^{-1}$ toward infinity.

Compared with the Einstein gravity produced by $G_{\text{zero}}$, which is proportional to $1/R$, the dynamical KK field seems to be suppressed in the region far from the isolated matter. 
However, the radial dependence of the metric for the KK modes is not necessarily suppressed compared to that of the Einstein gravity, as we will see explicitly below employing the areal radius for the spherically symmetric case. 
Thus in this sense, the dynamical KK modes could produce ``leading order'' contributions, and then it is interesting to study various effects induced by the dynamical KK corrections. 

Before closing this subsection, it is worthwhile to note that the above dynamical KK corrections could be interpreted into the context of conformal field theory via the AdS/CFT correspondence in the braneworld.
The correspondence has been ever discussed to investigate for example following aspects of the braneworld gravity;
$1/R^{3}$ correction to the Newtonian gravitational potential, homogeneous cosmology, and tensor-type perturbations on a FLRW brane~\cite{Duff,Ida, GKR, Tanaka04, KS, Gubser:2001}.
It is quite interesting to consider if this correspondence also persists in dynamical cases.

For dynamical sources ($\omega \neq 0$), the CFT one-loop correction $\Pi_2(p)$ to the graviton propagator~\cite{Duff} are reduced in the Fourier space to 
\begin{equation}
\begin{aligned}
\widehat \Pi_2(\omega;\mathbf x,0)
&=
\int \frac{\mathrm d^3 \mathbf k}{(2\pi)^3} \Pi_2(p) e^{i\mathbf
 k\cdot\mathbf x}
\simeq \frac{\ell^2}{8}\int \frac{\mathrm d^3 \mathbf k}{(2\pi)^3} e^{i\mathbf
 k\cdot\mathbf x}\ln (\mathbf k^2 - \omega^2)
\\
& = \frac{\ell^2}{32\pi^2 ir} \int_{-\infty}^\infty \!\!\!\mathrm d k e^{ikr}
 k\{\ln(k-\omega -i\epsilon) + \ln(k+\omega +i\epsilon)\}
\\
&= \frac{\ell^2}{16\pi}\left(\frac{i\omega}{r^2}-\frac{1}{r^3}\right)
 e^{i\omega r},
\end{aligned}
\end{equation}
where $\Pi_2(p)\simeq\frac{\ell^2}{8}\ln p^2$ and $p^2 =-(\omega +
i\epsilon)^2 + \mathbf k^2$.
This means that in dynamical cases the KK Green's function of
Eq.~(\ref{KK Green's function}) can be rewritten by the CFT correction $\Pi _2 (p^2)$.
Therefore it is possible that the dynamical external field discussing below could be interpreted via CFT description.

\subsection{Quadrupole formula for KK fields}

In this subsection we shall write the metric perturbation of the KK excitations in terms of quadrupole moment of matter distribution.
The quadrupole formula is very common in General Relativity. 
In 4D General Relativity corresponding to the zero mode, the metric perturbation is given under an appropriate gauge~\cite{MTW} by
\begin{equation}
\begin{aligned}
h^\mathrm{zero}_{00}(t,\mathbf x) =&
 \frac{2G_4M}{r} + \mathcal O(r^{-3}),\\
h^\mathrm{zero}_{0i}(t,\mathbf x) =& 
 \frac{G_4}{r^2}\frac{x_j}{r}J^{ij} + \mathcal O(r^{-3}),\\
h^\mathrm{zero}_{ij}(t,\mathbf x) =&
\frac{2G_4M}{r}\delta_{ij} +
 \frac{2G_4}{r}\frac{\mathrm d^2}{\mathrm d t^2}I^\mathrm{TT}_{ij}(t-r) +
 \mathcal O(r^{-3}),
\end{aligned}
\label{eq:4DGR quadrupole}
\end{equation}
where
$I^\mathrm{TT}_{ij}$ is transverse-traceless components of quadrupole
moment and
$J^{ij} \equiv \int\mathrm d^3\mathbf x (x^iT^{j0}-x^jT^{i0})$ is total angular momentum. 
Thus in these leading terms only the transverse-traceless components are time-dependent, which describe gravitational waves.  
The other terms describe total mass $M$ and total angular momentum $J_{ij}$, which are conserved quantities (time-independent).

Let us now return to the KK modes. 
From Eq.~(\ref{solution}) the metric perturbation of the KK excitations on the brane at $z=\ell$ is 
\begin{equation}
\begin{aligned}
{\widehat{\bar h}}^\mathrm{KK}_{\mu\nu}(\omega,\mathbf x) 
&\simeq - 2G_4\ell^2\int\mathrm
 d^3\mathbf x' e^{i\omega
 R}\left(\frac{i\omega}{R^2}-\frac{1}{R^3}\right)
\left(\widehat T_{\mu\nu}-\frac{1}{3}\eta_{\mu\nu}\widehat T\right)
 (\omega,\mathbf x').
\label{KK metric1}
\end{aligned}
\end{equation}
We note that $\bar h_{\mu\nu}$ is obtained by the gauge transformation to the
Gaussian normal coordinates, and the gauge transformation $\widehat\xi^5$ in the source term of Eq.~(\ref{solution}) does not contribute to the KK modes~\cite{Garriga:1999yh}.

To obtain the quadrupole representation of the metric perturbation, we rewrite the integral of the energy-momentum tensor in terms of quadrupole moment.
By energy conservation, $\partial_\mu T^{\mu\nu}=0$, we have the following relation~\cite{Wald},
\begin{equation}
\begin{aligned}
\int \mathrm d^3\mathbf x \widehat T^{ij} &= \int\mathrm d^3\mathbf x
 [\partial_k (\widehat T^{kj}x^i)-x^i\partial_k \widehat T^{kj}]\\
&= -i\omega \int \mathrm d^3\mathbf x x^i \widehat T^{0j}\\
&= -\frac{i\omega}{2} \int \mathrm d^3\mathbf x [\partial_k
 (\widehat T^{0k}x^ix^j) - x^ix^j\partial_k \widehat T^{0k}]\\
&= -\frac{\omega^2}{2}\int \mathrm d^3\mathbf x x^ix^j\widehat T^{00}
\equiv -\frac{\omega^2}{2}\widehat I^{ij}(\omega),
\end{aligned}
\end{equation}
where $\widehat I^{ij}$ is quadrupole moment and $i,j,k$ are spatial indexes.
At this point, we assume $\omega |\mathbf x'| \ll 1$ in the slow-motion approximation that the internal velocities of sources are small and thus the typical source radius is much smaller than the wavelength $1/\omega$.
At large distances from the source, i.e., $r\equiv|\mathbf x|\gg\mathbf
x'$, it is possible to make multipole expansions up to the order of quadrupole moment: 
\begin{equation}
\begin{aligned}
\int\mathrm d^3\mathbf x' e^{i\omega R}\left(\frac{i\omega}{R^2} -
					\frac{1}{R^3}\right)\widehat T^{ij}
\simeq& - \frac{1}{2} \left[\frac{(-i\omega)^3}{r^2} +
		      \frac{(-i\omega)^2}{r^3}\right]
\widehat I^{ij}(\omega)e^{i\omega r},\\
\int\mathrm d^3\mathbf x' e^{i\omega R}\left(\frac{i\omega}{R^2} -
					\frac{1}{R^3}\right)\widehat T^{0j}
\simeq& - \frac{1}{2} \left[\frac{(-i\omega)^3}{r^2} +
		      3\frac{(-i\omega)^2}{r^3} +
 3\frac{(-i\omega)}{r^4}\right] \frac{x_i}{r}\widehat I^{ij}(\omega)e^{i\omega
 r},
\\
\int\mathrm d^3\mathbf x' e^{i\omega R}\left(\frac{i\omega}{R^2} -
					\frac{1}{R^3}\right)\widehat T^{00}
\simeq
&
 -\frac{M}{r^3}2\pi\delta(\omega)
  - \frac{1}{2}\left\{\left[\frac{(-i\omega)^3}{3r^2} +
 2\frac{(-i\omega)^2}{r^3} + 5\frac{(-i\omega)}{r^4} +
 5\frac{1}{r^5}\right]
 \left(3\frac{x_ix_j}{r^2}-\delta_{ij}\right)\right.\\
&\left. + \left[\frac{(-i\omega)^3}{3r^2} +
		      \frac{(-i\omega)^2}{r^3} +
 2\frac{(-i\omega)}{r^4} + 2\frac{1}{r^5}\right]\delta_{ij}\right\}
 \widehat I^{ij}(\omega)e^{i\omega r}.\label{up_to_quadrupole}
\end{aligned}
\end{equation}
Here in the rest frame 
$\int \mathrm d^3\mathbf x T^{00} = M$ and
$\int \mathrm d^3\mathbf x T^{0i} = 0$,
where $M$ is the total mass of the system.
Substituting these into Eq.~(\ref{KK metric1}), we obtain the metric perturbation expressed by the quadrupole moment, the results in the real space are 
\begin{equation}
\begin{aligned}
\bar h^\mathrm{KK}_{00}(t,\mathbf x) =&
 \frac{4MG_4\ell^2}{3r^3}
 + \frac{2G_4\ell^2}{3}
\left[\frac{5}{6r^2}\frac{\mathrm d^3}{\mathrm d t^3} +
		      \frac{3}{2r^3}\frac{\mathrm d^2}{\mathrm d t^2} +
 \frac{2}{r^4}\frac{\mathrm d}{\mathrm d t} +
 2\frac{1}{r^5}\right]I(t-r)\\
& + 2G_4\ell^2\left[\frac{1}{3r^2}\frac{\mathrm d^3}{\mathrm d t^3}
 + \frac{2}{r^3}\frac{\mathrm d^2}{\mathrm d t^2} +
 \frac{5}{r^4}\frac{\mathrm d}{\mathrm d t} +
 \frac{5}{r^5}\right]
 \frac{x^ix^j}{r^2}\cbar I_{ij}(t-r),\\
\bar h^\mathrm{KK}_{0i}(t,\mathbf x) =& 
-G_4\ell^2 \left[\frac{1}{r^2}\frac{\mathrm d^3}{\mathrm d t^3} +
		      \frac{3}{r^3}\frac{\mathrm d^2}{\mathrm d t^2} +
 \frac{3}{r^4}\frac{\mathrm d}{\mathrm d t}\right]
 \frac{x^j}{r}I_{ij}(t-r),\\
\bar h^\mathrm{KK}_{ij}(t,\mathbf x) =&
\frac{2MG_4\ell^2}{3r^3}\delta_{ij} +
 \frac{G_4\ell^2}{3}\left[\frac{1}{3r^2}\frac{\mathrm d^3}{\mathrm d t^3} +
 \frac{1}{r^3}\frac{\mathrm d^2}{\mathrm d t^2} +
 \frac{2}{r^4}\frac{\mathrm d}{\mathrm d t} +
 \frac{2}{r^5}\right]I(t-r)\delta_{ij}\\
&+ G_4\ell^2 \left[\frac{1}{r^2}\frac{\mathrm d^3}{\mathrm d t^3} +
 \frac{1}{r^3}\frac{\mathrm d^2}{\mathrm d t^2}\right]
\cbar I_{ij}(t-r),\label{KK metric2}
\end{aligned}
\end{equation}
where $I=I^k{}_k$ and $\cbar I_{ij} = I_{ij} -
\frac{1}{3}I \delta_{ij}$ is the trace-free part whose components will vanish if the source is spherically symmetric.
The terms of $\mathcal O(1/r^3)$ which are independent of the quadrupole moments are static KK corrections. 
If the matter source is dynamical, then the dynamical terms of order $1/r^2$ arise.

The result obtained here provides a formula that describes the far
exterior gravitational field induced by general matter sources localized
on the brane.
It shows explicitly that in the braneworld even spherical sources can make the exterior gravitational field dynamical.  
(See \cite{BGM} for the case of homogeneous dust collapse.)

Although the linear perturbation is used to represent weak field approximation, it does not imply the lack of possibility to describe strong-gravity systems (e.g., a self-gravitational system).
Because for asymptotically flat spacetimes the gravitational field will become so weak to be dealt as linear perturbations at sufficiently far region, the linearized Einstein equation can be thought of as an ``exact'' equation if other nonlinear terms are taken as a source term of the equation~\cite{MTW, Thorne:1980}.
In other words, even if the interior is strongly gravitating system
which needs nonlinear treatment, its far exterior field can be described
in the same manner as for a weakly gravitating system except the
definition of total mass or quadrupole moment.
(See, e.g.,~\cite{Blanchet:2002av}.)

\section{Applications}
\label{sec:Applications}

\subsection{Spherically symmetric case}

Now we consider a spherically symmetric source, of which consequences
are non-trivial in the braneworld.
By virtue of spherical symmetry, the quadrupole moment becomes
\begin{equation}
I_{ij}(t) = \frac{1}{3}\delta_{ij}I(t), 
\quad
I(t) \equiv I^k{}_k = \int\mathrm d^3\mathbf x ~r^2 \rho, 
\end{equation}
where the trace-free part $\cbar I_{ij}$ vanishes. 
In the spherical case the spatial components of the metric perturbation
$\bar h^\mathrm{KK}_{ij}$ are proportional to $\delta_{ij}$ so that the
metric can be expressed in the isotropic coordinates. 
The expression becomes concise after being transformed into the area
coordinates by following gauge transformation 
\begin{equation}
\begin{aligned}
\xi_t &= -\frac{G_4\ell^2}{18}\left[\frac{1}{r}I^{(3)} +
 \frac{8}{r^2}I^{(2)} + \frac{8}{r^3}I^{(1)}\right],\\
\xi_r &= - \frac{MG_4\ell^2}{3r^2}
 -\frac{G_4\ell^2}{18}\left[\frac{1}{r}I^{(3)} + \frac{3}{r^2}I^{(2)} +
 \frac{6}{r^3}I^{(1)} + \frac{6}{r^4}I\right],
\end{aligned}
\end{equation}
where $\displaystyle I^{(n)} \equiv \frac{\mathrm d^n}{\mathrm d t^n}
I(t-r)$. 
The non-vanishing components in the area coordinates are 
\begin{equation}
\begin{aligned}
\tilde{\bar h}^\mathrm{KK}_{tt}(t,\mathbf x) =&
 \frac{4MG_4\ell^2}{3r^3}
 - \frac{G_4\ell^2}{9}
\left[\frac{1}{r}I^{(4)} + \frac{3}{r^2}I^{(3)} -
		      \frac{1}{r^3}I^{(2)} -
 \frac{12}{r^4}I^{(1)} -
 \frac{12}{r^5}I\right],\\
\tilde{\bar h}^\mathrm{KK}_{rr}(t,\mathbf x) =&
\frac{2MG_4\ell^2}{r^3} +
 \frac{G_4\ell^2}{9}\left[\frac{1}{r}I^{(4)} + \frac{5}{r^2}I^{(3)} +
 \frac{15}{r^3}I^{(2)} +
 \frac{30}{r^4}I^{(1)} +
 \frac{30}{r^5}I\right].\label{area coordinate}
\end{aligned}
\end{equation}
Compared with 4D General Relativity (\ref{eq:4DGR quadrupole}), a new
term proportional to $1/r$ appears in the radial component
$\tilde{\bar h}^\mathrm{KK}_{rr}$ in this coordinates which claims that
the KK modes may contribute to energy evaluated at infinity.

As mentioned above, the expression of the metric perturbation depends on
the gauge.
We show the contribution to the energy using the Einstein tensor which
is a gauge-invariant quantity in this order.
The effective 4D Einstein equation on a vacuum brane~\cite{Shiromizu:1999wj} is given by 
\begin{equation}
G_{\mu\nu} = -\mathcal E_{\mu\nu},
\label{vacuum_SMS}
\end{equation}
where $\mathcal E_{\mu\nu}$ comes from 5D Weyl tensor. 
Since now we have the induced metric on the brane, we can directory
obtain the effects from the bulk. 
Straightforward calculation of LHS in the linear order leads to 
\begin{equation}
\begin{aligned}
\mathcal E^t{}_t \simeq& - \frac{4MG_4\ell^2}{r^5} -
 G_4\ell^2\left[\frac{1}{9r^2}I^{(5)} + \frac{5}{9r^3}I^{(4)} +
 \frac{20}{9r^4}I^{(3)} +
 \frac{20}{3r^5}I^{(2)} + \frac{40}{3r^6}I^{(1)} +
 \frac{40}{3r^7}I\right],
\\
\mathcal E^r{}_r \simeq& - \frac{2MG_4\ell^2}{r^5} +
 G_4\ell^2\left[\frac{1}{9r^2}I^{(5)} + \frac{5}{9r^3}I^{(4)} +
 \frac{10}{9r^4}I^{(3)} -
 \frac{10}{3r^6}I^{(1)} -
 \frac{10}{3r^7}I\right],\\
\mathcal E^\theta{}_\theta =& \mathcal E^\phi{}_\phi \simeq \frac{3MG_4\ell^2}{r^5} + G_4\ell^2
 \left[\frac{5}{9r^4}I^{(3)} +
 \frac{10}{3r^5}I^{(2)} + \frac{25}{3r^6}I^{(1)} +
 \frac{25}{3r^7}I\right],
 \\
\mathcal E^r{}_t \simeq& -G_4\ell^2\left[\frac{1}{9r^2}I^{(5)} +
		      \frac{5}{9r^3}I^{(4)} +
 \frac{5}{3r^4}I^{(3)} +
 \frac{10}{3r^5}I^{(2)} + \frac{10}{3r^6}I^{(1)}\right].
\label{E_munu}
\end{aligned}
\end{equation}
It is explicitly shown that $\mathcal E_{\mu\nu}$ satisfies the local conservation law in the linear order,
$\partial_\mu \mathcal E^{\mu\nu} = 0,$ and the trace-free condition, $\mathcal E^\mu{}_\mu = 0$.
Since $\mathcal E_{\mu\nu}$ can be regarded as an {\it{effective}} energy-momentum tensor on the brane, we can discuss the effective energy flux of the KK modes corresponding to $\mathcal E_{\mu\nu}$.
The total energy flux of the KK modes, $L_\mathrm{KK}$, is evaluated at future null infinity on the brane ($r\rightarrow \infty, t-r=\text{const.}$) as 
\begin{equation}
L_\mathrm{KK}\equiv - \frac{\mathrm dE^\mathrm{KK}}{\mathrm dt} = -\frac{1}{8\pi G_4}
\frac{\mathrm d}{\mathrm dt} \int_V (-\mathcal E_{tt})\;\mathrm
d^3\mathbf x = -\frac{1}{8\pi G_4}\oint_{S_\infty} \!\!(-\mathcal
E_{ti})\;\mathrm dS^i = -\frac{\ell^2}{18}I^{(5)},\label{KK energy}
\end{equation} 
where $E^\mathrm{KK}$ is the KK part of the energy of the system.
Thus this is suppressed by $\ell^2$.
We should notice that Eq.~(\ref{KK energy}) is not quadratic in the quadrupole moment as contrasted with the well-known expression for the 4D gravitational radiation.
This is due to the fact that $\mathcal E_{\mu\nu}$ exists in the linear order of metric perturbation.
The sign of $\mathrm dE^\mathrm{KK}/\mathrm dt$ is indefinite, in other words, the KK modes do not always carry positive energy away in contrast to the zero mode \cite{Sasaki:1999mi}. We will show an example in the next subsection.

While been considering $\mathcal E_{\mu\nu}$ as the energy of the KK modes, in usual discussions of the energy of gravitational waves we use the energy-momentum pseudo-tensor of gravitational fields, which is quadratic in the metric perturbation~\cite{MTW,Wein}.
After a straightforward calculation of the Einstein tensor to the second order, ${}^{(2)}G_{\mu\nu}$, we find 
\begin{eqnarray}
{}^{(2)}G_{\mu\nu} \sim \mathcal O\left( \frac{1}{r^3} \right).
\end{eqnarray}
Thus the contributions are higher order, meaning that this kind of energy for the KK modes cannot reach infinity.
Consequently the KK modes carry its energy to null infinity on the brane not through the quadratic order (gravitational energy) but the linear order of
$\mathcal E_{\mu\nu}$, like an effective matter flux.
We will observe another aspect of this nature in Appendix, in which we relax the assumption of the slow motion and it turns out that
$\mathcal E_{\mu\nu}$ has a flux part explicitly.

\subsection{Spherical dust collapse}

As a concrete, simple example of dynamical systems we shall consider homogeneous dust collapse under spherical symmetry.
This model is well known as the Oppenheimer-Snyder model in the 4D case.
As our attention is concentrated on the weak gravitational field induced by non-relativistic matter, we deal with the matter source as Newtonian.
We denote the mass, radius, and density of the dust cloud by $M$, $R$, and $\rho$, respectively. 
The equation of motion of the dust sphere is given by
\begin{equation}
\left(\frac{\mathrm d R}{\mathrm d t}\right)^2 \simeq \frac{r_\mathrm S}{R},
 \quad \rho(t) = \frac{3M}{4\pi R^3(t)},
\end{equation}
where $r_\mathrm S \equiv 2G_4M$ is the 4D Schwarzschild radius. The quadrupole moment of this system is $I(t) = \frac{3}{5}MR^2(t) $.
Here we have neglected the KK corrections to the equation of motion and the time derivative of quadrupole moment, because they only give higher order corrections in the resulting metric $h^\mathrm{KK}_{\mu\nu}$.

Substituting these into Eqs.~(\ref{area coordinate}) and (\ref{KK energy}), we obtain
\begin{equation}
\tilde{\bar h}^\mathrm{KK}_{tt}(t,\mathbf x) =
 - 
\frac{\ell}{r}\frac{\ell}{12 R}\left(\frac{r_\mathrm
 S}{R}\right)^3 + \mathcal O(r^{-2}),
\end{equation}
and 
\begin{equation}
L_\mathrm{KK}=
-\frac{\ell^2}{6G_4R^2}\left(\frac{r_\mathrm S}{R}\right)^\frac{7}{2}.
\end{equation}
These results indicate that the exterior metric of the collapsing dust is manifestly non-static and the total flux of the KK modes is negative.
This means that the total effective energy of the system for an observer at infinity on the brane increases during the collapse.
This is the direct consequence of the negativity of the effective energy density of $\mathcal E_{\mu\nu}$. 
The gravitational fields are expressed using the retarded Green's function, so outside the sources any flux is outgoing, in other words, depends on retarded time $t-r$.
If the energy density of the flux is negative, the total energy on the brane will increase.
This kind of negative energy density for $\mathcal E_{\mu\nu}$ is general in the braneworld.
For example, the exact solution of the localized black hole in the 4D RS model~\cite{EHM} or the KK correction to the Newton potential for a static source~\cite{Garriga:1999yh} give the negative energy density.
Also in the brane cosmologies the KK modes behave as dust with a negative energy density on the brane~\cite{Minamitsuji:2005xs}.

\subsection{Non-spherically symmetric case}

We shall now briefly discuss non-spherical cases.
The terms including $\cbar I_{ij}$ remain in Eq.~(\ref{KK metric2}) and give additional terms for $ \mathcal E_{\mu\nu} $ other than those in
Eq.~(\ref{E_munu}).
The relevant component to such energy flux is $\mathcal E_{tr}$ which is
evaluated as
\begin{equation}
\mathcal E^\mathrm{non}_{tr}
 = - \frac{G_4\ell^2}{3r^2} \frac{x^ix^j}{r^2} \cbar I_{ij}^{(5)}
   + \mathcal O(r^{-3}),
\end{equation}
This leads to the energy change rate due to the non-spherical terms.
However, it turns out that the rate is zero,
\begin{equation}
\frac{\mathrm dE^\mathrm{out}}{\mathrm dt}
 \propto \oint_{S_\infty}\!\!
         \frac{x^ix^j}{r^2} \cbar I_{ij}^{(5)} \mathrm d\Omega
 = \frac{4\pi}{3}\delta_{ij}\cbar I_{ij}^{(5)}
 = 0,
\end{equation}
and that {\it the non-spherical part of the KK modes cannot carry the energy}.
This means that for general dynamical processes such as particles infalling into a black hole or binary stars, only the spherical part is relevant to the total flux of the KK modes.
This is remarkable contrast to the emission mechanism of ordinary gravitational waves (zero modes).

We shall compare the energy flux of the KK modes with that of the zero mode.  
The total energy flux of the zero mode is given by
\begin{equation}
L_\mathrm{zero}
 = \frac{G_4}{5}\cbar I_{ij}^{(3)}\cbar I_{ij}^{(3)},
\end{equation}
and we estimate
\begin{equation}
\left|\frac{L_\mathrm{KK}}{L_\mathrm{zero}}\right|
 \sim \frac{\ell}{G_4M} \frac{\ell}{R} \frac{(cT)}{R}
~ \sim ~
10^{-16}\times 
\left(\frac{\ell/0.1\text{mm}}{G_4M/M_\odot}\right)^{3/2}
\left(\frac{\ell/0.1\text{mm}}{R/10\text{km}}\right)^{1/2}
\end{equation}
where $M$,$R$ and $T$ are characteristic mass, source radius and
time-scale, respectively, taken to be
$\ell \ll R \lesssim cT \sim \sqrt{R^3/G_4M}$. 
Thus the relative amount of energy carried by the KK modes is negligible in general astrophysical situations.

\section{summary and discussion}
\label{sec:summary and discussion}

We have discussed the gravitational fields surrounding isolated dynamical systems in the RS model and obtained the quadrupole formula for the KK modes.  
We found that in dynamical cases the leading term of the KK correction for the Green's function is of order $1/r^2$ rather than $1/r^3$, which is observed in the static case.  
By applying the quadrupole formula, we have considered the effective energy flux of the KK modes.  
Contrary to the fact that ordinary gravitational waves (the zero mode) are generated by non-spherical parts of matter distributions, the effective energy flux of the KK modes arises only through the spherical part.  
This implies that it is difficult to generate the flux of the KK modes
by a circular motion such as binary pulsars.
Moreover, even for spherical collapses which cannot radiate the zero
mode, the radiated KK modes are suppressed by $\ell^2$ as well as static
cases~\cite{RS2,Garriga:1999yh}.

For spherical dust collapse we have explicitly derived the exterior metric on the brane and evaluated the effective energy flux during the collapse. 
In our treatment, the KK modes contribute to the total energy of systems
on the brane through the projected bulk Weyl tensor, 
$\mathcal E_{\mu\nu}$, but not through the energy-momentum tensor of gravitational field itself. 
This result is a consequence of the fact that in the 4D picture of the braneworld the KK modes induce the energy-momentum as an effective matter in the form of $ - \mathcal E_{\mu\nu}$.

In the linear order, $\mathcal E_{\mu\nu}$ satisfies the local conservation law as the energy-momentum tensor of matter fields satisfies, and hence it has independent conserved charges at this order.
In this sense the effective energy flux of the KK modes is irrelevant to the energy loss via gravitational wave radiation.
To derive the correction to the energy loss rate due to KK gravitons, we
will need to go to second-order perturbations~\cite{Kudoh:2001wb}(see
also, e.g.,~\cite{Blanchet:1986dk} for ordinary gravitational waves).

In nonlinear regime the KK modes couple to the matter distribution due to the relation $\nabla^\mu ({\mathcal{E}}_{\mu\nu} - 
\frac{48\pi G_4}{\lambda} \pi_{\mu\nu})=0$ where $\pi_{\mu\nu}$ consists of the energy-momentum tensor $T_{\mu\nu}$ \cite{Shiromizu:1999wj}, and it does not have a separate conserved charge.
Thus the next-order perturbations will affect the energy of matter,
and it will reveal us how the ordinary gravitational wave forms and the energy loss rate are affected by the KK modes. 
It is interesting to derive a quadrupole formula as a result of the second-order
interaction and we expect to provide new tests for the braneworld. 
The analysis of the nonlinear interaction and energy loss rate will be
discussed in a forthcoming paper~\cite{Kinoshita}.
(A brief discussion is given in Ref. \cite{Inoue:2003di})

The flux of the KK modes works as an effective matter flux,
namely a flux of the dark radiation.
This type of energy flux is not necessarily to be positive.
In fact, as we have observed in the case of the dust collapse, the KK modes have negative energy density.
As a result the total charge associated with $ \mathcal{E^{\mu\nu}}$ increases after the collapse.
This is related to the fact that the 5D Weyl tensor $\mathcal  E_{\mu\nu}$ in the effective Einstein equations appears in the linear order of perturbations so that the sign of $\mathcal E_{\mu\nu}$ depends on that of the metric perturbation.

We expect that the results can be understood in terms of the AdS/CFT correspondence because it was confirmed that the KK Green's function for dynamical processes corresponds to the CFT correction.
It is interesting to give physical meaning to the negative energy flow via the CFT picture ~\cite{Inoue:2003di}.

\begin{acknowledgments}  
We gratefully acknowledge helpful discussions with Takahiro Tanaka.
We would like to thank T. Shiromizu and K.I. Maeda for their valuable
 comments.
This work was partially supported by Grant-in-Aid for the Scientific
 Research from the Ministry of Education, Culture, Sports, Science and
 Technology of Japan through No.~14102004. 
H.K. and Y.S. are supported by JSPS.
\end{acknowledgments}    

\appendix  
\section{beyond slow-motion approximation}
In this appendix we relax the assumption of the slowly moving matter
sources used throughout this paper.
Then one can argue the gravitational wave from, for example, head-on
collision of relativistic particles~\cite{Wein, Lemos}.

Let us first recall our treatment within the slow-motion approximation
 ($\omega |\mathbf x'|\ll 1$).
We could write the integral in Eq.~(\ref{up_to_quadrupole}) in term of the quadrupole moments as
\begin{equation}
\int\mathrm d^3\mathbf x' e^{i\omega |\mathbf x - \mathbf x'|}
T^{ij}(\omega, \mathbf x') \simeq 
e^{i\omega r}\int\mathrm d^3\mathbf x' 
T^{ij}(\omega, \mathbf x') = 
-\frac{\omega^2}{2}I^{ij}(\omega)e^{i\omega r},
\end{equation}
where $|\mathbf x'| \ll |\mathbf x| \equiv r$ and we set
$e^{-i\omega|\mathbf x'|} \sim 1$.
However, in more general cases, we cannot neglect the phase factor
$e^{-i\omega |\mathbf x'|}$, and the above integral is estimated as 
\begin{equation}
\begin{aligned}
\int\mathrm d^3\mathbf x' e^{i\omega |\mathbf x - \mathbf x'|}
T^{\mu\nu}(\omega, \mathbf x') \simeq& 
e^{i\omega r}\int\mathrm d^3\mathbf x' 
e^{-i\omega \frac{\mathbf x\cdot\mathbf x'}{|\mathbf x|}}
T^{\mu\nu}(\omega, \mathbf x')\\
 =& 
e^{i\omega r}\int\mathrm d^3\mathbf x' 
e^{-i\mathbf k\cdot\mathbf x'}
T^{\mu\nu}(\omega, \mathbf x')
= \widehat T^{\mu\nu}(\omega,\mathbf k)e^{i\omega r},
\end{aligned}
\end{equation}
where $\mathbf k \equiv \omega \mathbf x /|\mathbf x|$ and the last
equality is the Fourier transformation with respect to spatial
coordinates.
Note that the conservation law yields $k_\mu \widehat T^{\mu\nu} = 0$ where $k^\mu = (\omega, \mathbf k)$ and $k^\mu k_\mu = 0$.

From Eq.~(\ref{KK metric1}), the leading order of the KK modes without the slow-motion approximation becomes
\begin{equation}
\begin{aligned}
h_{\mu\nu}^\mathrm{KK} =& -2\kappa \int\mathrm d^3\mathbf x'G_\mathrm{KK}
\left(T_{\mu\nu} - \frac{1}{3}\eta_{\mu\nu} T\right)(\omega, \mathbf x')\\
\simeq& -2G_4\ell^2 \frac{i\omega}{r^2}e^{i\omega r}
\left(\widehat T_{\mu\nu} - \frac{1}{3}\eta_{\mu\nu} \widehat
 T\right)(\omega, \mathbf k)
+ \mathcal O(r^{-3}),
\end{aligned}
\end{equation} 
and the corresponding projected Weyl tensor is 
\begin{equation}
\mathcal E_{\mu\nu} = \frac{G_4\ell^2}{3}\frac{i\omega}{r^2}
\widehat T k_\mu k_\nu + \mathcal O(r^{-3}).
\end{equation}
This explicitly indicates that the leading term of $\mathcal E_{\mu\nu}$ behaves as a null fluid.


\end{document}